\documentclass{mem}

\usepackage{natbib}
\usepackage{txfonts}
\usepackage{balance}
\usepackage{graphicx}
\usepackage[a4paper]{hyperref}
\idline{00}{45}

\newcommand{\kms}{km\,s$^{-1}$\,}
\newcommand{\pattern}{km\,s$^{-1}$\,kpc$^{-1}$\,}

\begin{document}

\title{Deriving the pattern speed using dynamical modelling of 
  gas flows in barred galaxies }

\subtitle{}

\author{I. \,P\'erez\inst{1,2} \and K. C. \, Freeman\inst{3}
 \and R. \, Fux\inst{4} \and A. \, Zurita\inst{2} }

\offprints{Isabel P\'erez; \email{isa@astro.rug.nl}}
 
\institute{Kapteyn Astronomical Institute, University of Groningen,
  Groningen, The Netherlands
\and
  Departamento de F\'isica te\'orica y del Cosmos, Universidad de
  Granada, Granada, Spain
\and
  Research School of Astronomy and Astrophysics, Australian National
  University, Weston Creek, ACT, Australia
\and
  Observatoire de Gen\`eve, Universit\`e de Gen\`eve, Sauverny,
  Switzerland }

\authorrunning{Perez et al.}
 
\titlerunning{Pattern speeds from dynamical modelling}

\abstract{In this paper we analyse the methodology to derive the bar
  pattern speed from dynamical simulations. The results are robust to
  the changes in the vertical-scale height and in the mass-to-light
  ($M/L$) ratios. There is a small range of parameters for which the
  kinematics can be fitted. We have also taken into account the use of
  different type of dynamical modelling and the effect of using 2-D vs
  1-D models in deriving the pattern speeds. We conclude that the
  derivation of the bar streaming motions and strength and position of
  shocks is not greatly affected by the fluid dynamical model used. We
  show new results on the derivation of the pattern speed for
  NGC~1530. The best fit pattern speed is around 10~\pattern, which
  corresponds to a $R_{\rm cor}/R_{\rm bar} = 1.4$, implying a slower
  bar than previously derived from more indirect assumptions. With
  this pattern speed, the global and most local kinematic features are
  beautifully reproduced. However, the simulations fail to reproduce
  the velocity gradients close to some bright HII regions in the
  bar. We have shown from the study of the H${\alpha}$ equivalent
  widths that the HII regions that are located further away from the
  bar dust-lane in its leading side, downstream from the main bar
  dust-lane, are older than the rest by 1.5--2.5\,Myr. In addition, a
  clear spatial correlation was found between the location of HII
  regions, dust spurs on the trailing side of the bar dust-lane, and
  the loci of maximum velocity gradients parallel to the bar major
  axis.
\keywords{galaxies: kinematics and dynamics -- galaxies: spiral 
  -- galaxies: structure} }

\maketitle{}

\section{Introduction}

The speed at which the bar rotates is one of the fundamental
ingredients describing bars, their evolution and their coupling with
other galaxy components. Its importance is indeed reflected in the
fact that this whole volume is devoted to the study of pattern speeds.

Some conclusions were already stated in the 90's about bar pattern
speeds (Elmegreen 1996). The pattern speed seems to be close to the
angular circular velocity in the disk near the end of the bar. The
corotation radius seems to be around 1.2 times the bar semi-major
axis, at least for early type galaxies. The question about the
dependence of the pattern speed with morphological type was also
raised and it is still an open question subject to discussion
(Rautiainen et al. 2005). It was also already concluded that the best
method for obtaining the location of corotation was that proposed by
Tremaine \& Weinberg (1984), hereafter TW method) which uses a set of
simple kinematic measurements to derive the pattern speed assuming
that the tracer obeys the continuity equation, that the disks are flat
and that there is one well defined pattern speed. Some of these
assumptions have been recently challenged, there is now a simple
extension of the TW method to multiple pattern speeds (Maciejewski
2006) and the fact that Beckman et al. (this volume) have shown that
the TW method can be applied to H${\alpha}$ velocity fields raises
some interesting issues.

Some indirect methods to derive the bar pattern speed include
identifying morphological or kinematic features with resonances (e.g.,
Athanassoula 1992, shape of dust lanes; Canzian 1993, sign inversion
of the radial streaming motion across corotation; Buta 1986, 1995,
rings as resonance indicators; Zhang \& Buta 2007, phase-shift between
the potential and density wave patterns). Other methods are based on
numerical modelling: (1) matching numerical simulations to the
observed velocity fields, from a set of general self-consistent models
or models deriving the potential from the light distribution (e.g.,
Duval \& Athanassoula 1983; Lindblad et al. 1996; Weiner et al. 2001;
P\'erez et al. 2004; Z\'anmar-S\'anchez et al. 2008; P\'erez \&
Zurita, in prep.), and (2) numerical simulations matched to galaxy
morphology (e.g., Hunter et al. 1988; England 1989; Rautiainen et
al. 2005).  The best indirect method, so far, to calculate pattern
speeds is the method based on the comparison of gas velocities to
those obtained in numerical simulations that use a potential obtained
from optical or NIR light (Elmegreen 1996).

In this paper, we will investigate the derivation of pattern speeds
from dynamical simulations, and we will analyse the possible problems
and caveats that one encounters when deriving bar pattern speeds in
this indirect way. Finally, we will present some new results on the
bar of NGC~1530.

\begin{figure*}
\resizebox{\textwidth}{!}{\includegraphics{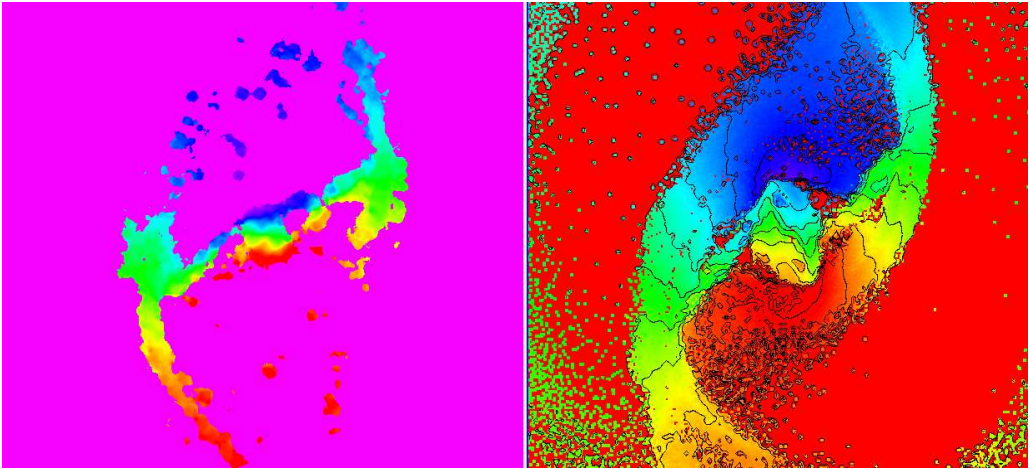}}
\caption{\footnotesize
  The left panel shows the velocity field from the H${\alpha}$
  observations by Zurita et al. (2004). The right panel shows the
  modelled velocity field with $R_{\rm cor}/R_{\rm bar}=1.4$. The
  velocity ranges between $+200$ and $-200$\,\kms }
\label{li_vhel}
\end{figure*}

\section{ Deriving the pattern speed}

The method to derive the pattern speeds from dynamical modelling is
relatively simple. From an optical or NIR image one can derive the
gravitational potential with some assumptions for the $M/L$ ratio and
the vertical scale-height, and imposing a certain pattern speed, then
gas flows are modelled using a fluid dynamical method, exploring a
certain range in parameter space. The modelled gas dynamics is then
compared to the observed kinematics until the best-fit solution is
found.

The N-body and hydro code, used by us in this work, was initially
developed by the Geneva Observatory galactic dynamics group for spiral
galaxy studies (Fux 1999, 1997; Pfenniger \& Friedli 1993). The
initially self-consistent code was modified to use a fixed rotating
potential. The stellar potential is fixed using the observed light
distribution. For more details about the code, refer to Fux (1999) or
his Ph.D. Thesis, Fux (1997). For more details about the code in the
form used for this work refer to P\'erez et al. (2004).

In this way, we have modelled 5 barred galaxies (NGC~5505, NGC~7483,
NGC~5728, and NGC~7267). Other groups that have taken the fluid
dynamical modelling approach to derive the bar dynamics, using the
light distribution to derive the potential, are Weiner et al. 2001,
modelling NGC~4123 and Z\'anmar-S\'anchez et al. (2008) modelling
NGC~1365, both using a 2-D Eulerian grid code. All these studies
concluded that the corotation radius is close to the end of the bar
(i.e., fast bars).

Now, we will briefly analyse the effect of changing the model
parameters on the output pattern speed.

The calculation of the $M/L$ from the broad--band colors using
population synthesis models is robust to most of the parameters going
into the modelling (Bell \& de Jong 2001; P\'erez 2003), such as SFR,
although, the absolute normalisation depends on the chosen IMF. The
photometric bands for which the $M/L$ ratio is most robust to galaxy
colour changes are the NIR bands.

As already shown in P\'erez et al. (2004), the scale height and the
$M/L$ ratio affect the location and amplitude of the shocks in the bar
region. For the vertical distribution an exponential profile is
assumed with a radially constant vertical scale-height. This may not
be a good assumption as there is evidence from infrared photometric
studies of the Milky Way that the vertical scale of the bar is larger
than that of the disk (Freudenreich 1998). The biggest impact on the
dynamics of a non-constant scale-height is precisely in the inner
region we are interested in, where the radial forces will change
significantly. However, it is hard to do better than this since not
much is known about the scale-heights of bars in external
galaxies. For most of the simulations one value of the scale-height
was adopted, following the relationship found by Kregel et al. (2002).
They analysed the structure of the stellar disk in a sample of edge-on
galaxies and found that the average $\langle h_R/h_z \rangle =7.3 \pm
2.2$, where $h_R$ and $h_z$ are respectively the disk exponential
scale-length and scale-height.  For the scale-heights smaller than
$h_z$, as defined before, none of the pattern speeds gives a good fit
for the studied galaxies. For the models with $h_{z} \times 1.5$ the
best fit corresponds to a slower bar; however, an average of
$\approx1$\,kpc is an unrealistically large value for the vertical
scale of a real galaxy with the observed scale-length (within the
ranges of scale-lengths analysed).

Varying the $M/L$ for a given pattern speed causes a variation in the
positions and strength of the shocks. The derivation of the pattern
speed is robust to $M/L$ changes; i.e., there is a small range of
$M/L$ ratios for which the models fit the observed kinematics (Weiner
et al. 2001; P\'erez et al. 2004). If we derive the models keeping the
$M/L$ ratios and varying the pattern speeds, the shock gets steeper
for slower bars and not only the morphology changes, but also the gas
flow changes and the locations and amplitude of the shocks no longer
fits the kinematics. The pattern speed is constrained within 10--20\%.
  
We have also taken into account the use of different type of dynamical
modelling and the effect of using 2-D vs 1-D in deriving the pattern
speeds. P\'erez (2008) showed that the global velocity field and the
gas distribution is very similar in both 2-D and 3-D models. The study
shows that the position and strength of the shocks developed in the
3-D N-body SPH simulations do not vary significantly compared to the
results derived from the 2-D Eulerian code. The results obtained in
the studies deriving the barred galaxies using the bar streaming
motions and strength and position of shocks are robust to the fluid
dynamical model used. Therefore, the effect of 2-D and 3-D modelling
can be neglected in this type of studies.
 
Summarising, the pattern speed is constrained within 10--20\% using
this indirect approach. All the galaxies studied in this way by the
different groups give pattern speeds that locate corotation close to
the end of the bar. The morphological types covered for these type of
simulations include from SBa to SBc galaxies, however, only a small
number of galaxies have been modelled in this way.

\section{Modelling NGC 1530}

NGC~1530 is an archetypal barred galaxy, at intermediate
inclination (between 40$^{\circ}$--50$^{\circ}$ according to different
studies), it has one of the largest bars ever observed, around 24 kpc
in length. The bar is dominated morphologically in the optical bands by
dust lanes running straight along the bar. It has a complicated
morphology in the nuclear region characterised in the optical and CO
bands by a nuclear ring/spiral.  NGC~1530 has been classified as a
class 7 torque (Block et al. 2004) using the dust-penetrated
classification scheme of Block \& Puerari (1999), this makes NGC~1530
one of the strongest bars found in the literature.  The weak inner
pseudoring in this object is made from arms that break from the ends
of the bar and wrap around the other ends (Block et
al. 2004). Although it has been widely studied, there has not been an
attempt to specifically model in detail the velocity field of NGC~1530
and, so far, only general SPH models of barred galaxies have been used
to compare the global trends of the observed velocity field of
NGC~1530.  Due to the richness of the bar region structures, and the
availability of high $S/N$ H${\alpha}$ velocity field (Zurita et al.
2004) NGC~1530 is a perfect candidate to test whether N-body/SPH can
reproduce the details of this galaxy's morphology and velocity field
and whether this type of simulations can teach us something else about
the dynamics of this galaxy.

Indirect assumptions have lead to the different authors to place
corotation at 1.0 and 1.2 times the bar semi-major axis, with derived
pattern speeds of 12 and 20~\pattern, respectively (Downes et
al. 1996; Regan et al. 1996), either assuming where the ILR resonance
or where corotation are located, and always based on the fact that the
size of the nuclear spiral/ring is around 1.5\,kpc in radius, and
corresponds to the location of the inner Lindblad resonance. We have
carried out a detailed modelling of NGC~1530 to study the position of
the resonances, and the correspondence of the dynamics with the star
formation in this galaxy. As already mentioned, this is not the first
attempt to understand the relation between the star formation and the
morphology and kinematics of this galaxy but this is the first attempt
to do a detailed comparison with numerical models, where the potential
is derived directly from the light distribution of NGC~1530.

\begin{figure*}
\resizebox{\textwidth}{!}{\includegraphics{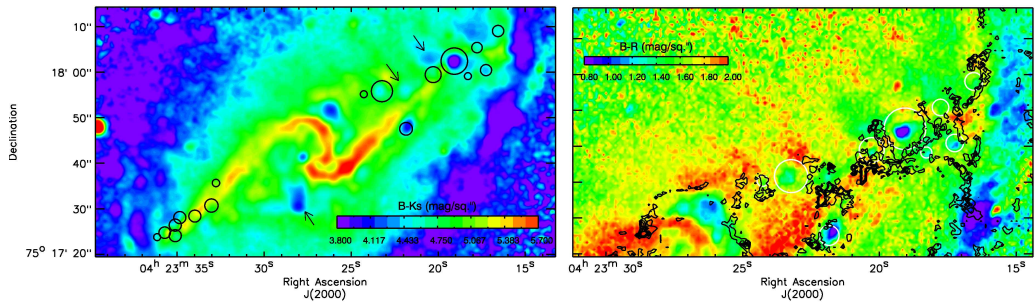}}
\caption{\footnotesize 
  Left panel shows the location of the studied HII regions on top of
  the $V$-$R$ color map, on the right panel a zoom on the north part
  of bar region with the observed velocity gradients
  over-plotted. Notice the correspondence between the dust spurs
  velocity gradients parallel to the bar major axis (Zurita \& P\'erez
  2008).}
\end{figure*}

The best fit pattern speed is around 10~\pattern, which corresponds to
a $R_{\rm cor}/R_{\rm bar} = 1.4$, which implies a slower bar than
previously derived from more indirect assumptions. With this pattern
speed the global kinematic features are beautifully reproduced
(Fig.~1). The models also develop the inner spiral/ring at around
1.2\,kpc that in previous studies was associated to the first ILR. For
this pattern speed, we find two ILRs one at around 0.15\,kpc and the
other at 4\,kpc. The results are compatible with a maximum disk. The
presence of an inner ring/spiral necessarily implies the existence of
an ILR, as previously suggested; however, the radius of the
ring/spiral need not be, necessarily, at the location of the resonance
(P\'erez \& Zurita, in prep.). We also find that the SPH models fail
to fully reproduce the exact location of the velocity gradients
observed along the bar, on the north side, close to some bright HII
regions. The HII regions in the bar region have been studied in detail
in Zurita \& P\'erez (2008). They concluded from the study of the
H${\alpha}$ equivalent widths that the HII regions that are located
further away from the bar dust-lane in its leading side, downstream
from the main bar dust-lane, are older than the rest by
~1.5--2.5\,Myr. In addition, a clear spatial correlation was found
between the location of HII regions, dust spurs on the trailing side of
the bar dust-lane, and the loci of maximum velocity gradients parallel
to the bar major axis (possibly tracing gas flow towards the main bar
dust-lane), see Fig.~2. These results support the hypothesis that
massive stars are forming on the trailing side of the bar dust-lane,
and age as they cross the bar, on a timescale that is compatible with
the bar dynamics time--scales.

\section{Conclusions}
 
We have analysed the use of dynamical simulations deriving the
potential from the galaxy light distribution to obtain the best-fit
pattern speed when compared to observed kinematics. There is a small
range of parameters for which the kinematics can be fitted. We have
analysed the effect of varying these parameters, namely the $M/L$
ratio, the pattern speed and the vertical scale-height. We can
constrain the pattern speeds within 10--20\%. All the galaxies
modelled in this way give pattern speeds that place corotation near
the end of the bar. The results are also unaffected by the choice of
the model in the dynamical modelling (P\'erez 2008).  We have,
therefore, refined the methods (e.g., $M/L$ derivation) and the study
of the systematics in this type of modelling but only for a handful of
galaxies have the gas flows been derived (Z\'anmar-S\'anchez et
al. 2008; P\'erez et al. 2004; Weiner et al 2001). This way of
deriving the pattern speed is perfect to study later types which
present nebular emission in the bar region. We should also compare, in
the future, these results with sticky--particle simulations to check
the reliability of the results, and improve the sample, carrying out
the modelling in overlapping samples where the pattern speeds have
been derived using different methods (cf., TW, morphology, etc.).

Unfortunately, we have not fully addressed yet with this method the
study of later morphological types. We have presented here first
results on the detail comparison of the modelled dynamics with the 2-D
H${\alpha}$ velocity fields for NGC~1530. The models reproduce the
global and local velocity fields (detailed results in P\'erez \&
Zurita, in prep.). However, it fails to reproduce the velocity
gradient on the north-side of the bar close to a few bright HII
regions. Analysis of the H${\alpha}$ equivalent widths show that the
HII regions that are located further away from the bar dust-lane in
its leading side, downstream from the main bar dust-lane, are older
than the rest by 1.5--2.5\,Myr. In addition, a clear spatial
correlation was found between the location of HII regions, dust spurs
on the trailing side of the bar dust-lane, and the loci of maximum
velocity gradients parallel to the bar major axis (Zurita \& P\'erez
2008).
 
\begin{acknowledgements} 
I. P\'erez is supported by a postdoctoral fellowship from the
Netherlands Organisation for Scientific Research (NWO, Veni-Grant
639.041.511) and the Spanish Plan Nacional del Espacio del Ministerio
de Educaci\'on y Ciencia. A.Z acknowledges support from the
Consejer\'{\i}a de Eduaci\'{o}n y C iencia de la Junta de
Andaluc\'{\i}a.
\end{acknowledgements}

%%% BIBLIOGRAPHY

\bibliographystyle{aa}

\end{document}